\documentclass[10 pt, conference]{IEEEtran}

\usepackage{amsmath,amssymb}
\usepackage{epsfig}

\newtheorem{theorem}{Theorem}
\newtheorem{example}{Example}
\newtheorem{remark}{Remark}

\newcommand{\te}{T_{\epsilon}^{(n)}}

\newcommand{\bb}{\mathbf}
\newcommand{\nn}{\nonumber}

\newcommand{\mc}{\mathcal}

\newcommand{\qed}{\nobreak \ifvmode \relax \else
      \ifdim\lastskip<1.5em \hskip-\lastskip
      \hskip1.5em plus0em minus0.5em \fi \nobreak
      \vrule height0.75em width0.5em depth0.25em\fi}

\begin{document}

\title{Deterministic Relay Networks with State Information}

\author{
\authorblockN{Sung Hoon Lim}
\authorblockA{School of EECS\\ KAIST\\ Daejeon, Korea \\
Email: sunghlim@kaist.ac.kr} \and
\authorblockN{Young-Han Kim}
\authorblockA{Department of ECE\\
UCSD\\
La Jolla, CA 92093, USA\\
Email: yhk@ucsd.edu } \and
\authorblockN{Sae-Young Chung}
\authorblockA{School of EECS\\ KAIST\\ Daejeon, Korea\\
Email: sychung@ee.kaist.ac.kr } }

\maketitle

\begin{abstract}
 Motivated by fading channels and erasure channels,
the problem of reliable communication over deterministic relay
networks is studied, in which relay nodes receive a function of the
incoming signals and a random network state. An achievable rate is
characterized for the case in which destination nodes have full
knowledge of the state information. If the relay nodes receive a
{\em linear} function of the incoming signals and the state in a
finite field, then the achievable rate is shown to be optimal,
meeting the cut-set upper bound on the capacity. This result
generalizes on a unified framework the work of Avestimehr, Diggavi,
and Tse on the deterministic networks {\em with state dependency},
the work of Dana, Gowaikar, Palanki, Hassibi, and Effros on linear
erasure networks \emph{with interference}, and the work of Smith and
Vishwanath on linear erasure networks \emph{with broadcast}.
\end{abstract}

\section{Introduction}
 In their celebrated paper \cite{Ahlswede00} that opened the field of
network coding, Ahlswede \emph{et al.} found the multicast capacity
of wireline networks.  For wireless networks, however, there are
some new challenges for reliable communication compared to the
wireline network. Among them are \emph{broadcast} and
\emph{interference}, and there has been some
 work that deals with these two features. In
\cite{Ratnakar06}, the multicast capacity was shown for networks
that have deterministic channels with broadcast, but without
interference at the receivers. Deterministic networks were further
studied in \cite{Avestimehr07} to incorporate interference at the
receiving nodes, where the capacity for linear finite field networks
was found. These rather simple models were shown to give good
insights in solving real-world network problems. For example in
\cite{Avestimehr08}, Avestimehr \emph{et al.} were able to
approximately characterize the capacity of Gaussian relay networks
within some constant gap using a similar approach used for
deterministic networks. Although previous models consider broadcast
and interference, they did not explicitly consider another important
feature in wireless communications. The wireless medium in
real-world communications suffer \emph{fading}, which in turn cause
severe degradation of the transmitted signal. Although the
deterministic model can be a good abstraction in understanding
broadcast and interference, it does not fully capture the effect of
fading in wireless networks. In this sense, the erasure network in
which transmitted symbols get erased at random provides a simple
model that captures the fading characteristics. In \cite{Dana06},
Dana \emph{et al.} considered the erasure networks with broadcast
and no interference, where the erasures are at the traversing edges.
Smith and Vishwanath \cite{Smith07} considered an erasure network
without broadcast, where the interference is modeled as a linear
finite field sum of incoming signals that are not erased. In both
\cite{Dana06} and \cite{Smith07}, if the destination node has
perfect knowledge of the state information, they showed that the
capacity is given by the cut-set bound.

 In this paper, we consider a deterministic network in which the observation at each node
is a function of the incoming signals and a random state. The
channel state affecting the relay and destination nodes is assumed
to be perfectly
known at the destinations. 
We give an achievable rate for this class of networks, and show that
the associated coding scheme achieves the capacity for the case in
which the relay and destination nodes receive a linear function of
the incoming signals and the state over a finite field. This result
generalizes the work of Dana \emph{et al.} and the work of Smith and
Vishwanath on linear erasure networks to handle both interference
and broadcast. As for deterministic networks, our result generalizes
the work of Avestimehr, Diggavi, and Tse on the deterministic
networks to deterministic state-dependent networks.

\section{Problem Statement and Preliminaries} \label{SEC:  Channel model}
 In the following we will give useful definitions for later use. Upper case letters denote
random variables (e.g., $X, Y, S$) and lower case letters represent
scalars (e.g., $x, y, s$). Calligraphic letters (e.g., $\mc A$)
denote sets and the cardinality of the set is denoted by $|\mc A|$.
Subscripts are used to specify node and time indicies. For example,
$X_u$ and $X_{u,i}$ denotes the signal sent at node $u$ and the
signal sent at node $u$ at time $i$, respectively. To represent a
sequence of random variables we use the notation
$X_v^n=X_{v,1},\ldots, X_{v,n}$. We will frequently use random
variables subscripted by sets to denote the set of random variables
indexed with elements in the set. For example, $X_{\mc A}=\{X_a:
a\in \mc A\}$ and $X^n_{\mc A}=\{X^n_a: a\in \mc A\}$.

 We consider a network $\mc G=(\mc V, \mc E)$ where $\mc V$ and $\mc E$
are the set of nodes and directed edges, respectively. Without loss
of generality, we let $\mc V=\{1,\ldots, |\mc V|\}$ and index the
source node with $1$. We use $\mc D$ and $\mc R=\mc V-(\{1\}\cup \mc
D)$ to denote the set of destination nodes and relay nodes
respectively. The network has one channel input $X_u\in \mc X_u$
associated with each node $u\in \mc V$, where $\mc X_u$ is the
alphabet of $X_u$. This incorporates the broadcast nature of the
network. Each node $v\in \mc V$ observes
\begin{align}
Y_{v}=f_{v}\left(X_{\mc N_v}, S\right), \label{EQ: channel}
\end{align}
where the input neighbors $\mc N_v$ of $v$ is defined as $\mc N_v =
\{u : (u, v) \in \mc E\}$. The random variable $S$ is a random state
affecting nodes, which is independent of the source message. The
state sequence is memoryless and stationary with
$p(s^n)=\prod_{i=1}^n p(s_{i})$. We assume that each destination
$d\in \mc D$ has side information of the state sequence. The source
node wishes to send a common message $m\in [2^{nR}]\triangleq
\{1,\ldots,2^{nR}\}$ to all destination nodes.

 A $(2^{nR}, n)$ code consists of a source encoding function $\phi_1$,
relay encoding functions $\phi_{v,i}$, $v\in \mc V-(\{1\}\cup \mc
D)$, $i\in\{1,\ldots,n\}$, and decoding functions $\psi_d$, $d\in
\mc D$, where
\begin{align*}
&\phi_1: [2^{nR}] \rightarrow \mc X^n_1, \\
&\phi_{v,i}: \mc Y_v^{i-1} \rightarrow \mc X_{v},  i\in\{1,\ldots, n\}, v\in \mc R,\\
&\psi_d: \mc Y^n_d \times \mc S^n \rightarrow [2^{nR}], ~~ d\in \mc
D
\end{align*}
where $M$ is uniformly distributed over $[2^{nR}]$. The probability
of error is defined by
\begin{align*}
P_e^{(n)}=\text{Pr}\{\psi_d(Y_d^n, S^n)\neq M \text{ for some } d\in
\mc D\}.
\end{align*}
 A rate $R$ is said to be \emph{achievable} if there exist a sequence of $(2^{nR}, n)$ codes with $P_e^{(n)} \rightarrow 0$
as $n \rightarrow \infty$.

 For each $d\in \mc D$, a cut $\mc U_d \subset \mc V$ is a subset of nodes such that
$1\in \mc U_d$ and $d\in \mc U_d^c$. We will omit the destination
index when it is clear from the context. We define a boundary of a
cut as $\partial (\mc U)=\{u:(u,v)\in \mc E, u\in \mc U, v\in \mc
U^c\}$ and the boundary of a complement of a cut as $\bar{\partial}
(\mc U^c)=\{v:(u,v)\in \mc E, u\in \mc U, v\in \mc U^c\}$.

 We say that a node $v$ is in \emph{layer} $l$ if all
directed paths from the source to $v$ has $l$ hops. Let $L$ be the
longest distance from the source node to any node. We say that a
\emph{network is layered} with $L$ layers if every node in $\mc V$
belong to some layer $l\in\{0,\ldots,L\}$. The set of nodes in layer
$l$ is denoted by $\mc V_l$. Without loss of generality we will
assume that $\mc V_{0}=\{1\}$.

 For a random variable $X\sim p(x)$, the set $\te$ of $\epsilon$-typical
$n$-sequences $x^n$ is defined \cite{Orlitsky01} as
\begin{align*}
\te\triangleq \left\{x^n: \left|\pi(a|x^n)-p(a)\right|\leq
\delta\cdot p(a), \forall a\in \mc X\right\}
\end{align*}
where $\pi(a|x^n)$ is the relative frequency of the symbol $a$ in
the sequence $x^n$.
\section{Main result}
\subsection{General state dependent networks}\label{SEC: SD networks}

Given a class of relay networks as defined in (\ref{EQ: channel}),
the multicast capacity $C$ is upper bounded by
\begin{align}
C&\leq \max_{p(x_{\mc V})}\min_{d\in \mc D} \min_{\mc U_d}H( Y_{\mc
U_d^c}|X_{\mc U_d^c}, S). \label{EQ:cutset}
\end{align}
 The upper bound is from the cut-set bound \cite[Theorem 15.10.1]{Cover06}
by treating the state information as additional outputs to the
destinations, and using the fact that the state sequences are
independent of the message, the memoryless property of the channel,
and the deterministic nature of the channel given $S$.
\begin{remark} \label{RM: obliviousRelays}
The cut-set bound is given by (\ref{EQ:cutset}) whether we assume
that the relay nodes have state information or not, as long as the
state information at the relays are causal (i.e., $x_{v,i} =
\phi_{v,i}(y_v^{i-1},
   s^i)$)
 and destination nodes
have the state information.
\end{remark}
 As our main result we state the following theorem.
\begin{theorem} \label{TH: AchievableRate}
For the multicast relay network $\mc G=(\mc V, \mc E)$ in (\ref{EQ:
channel}), if all destination nodes in $\mc D$ have side information
of the state, then the capacity $C$ of the network is lower bounded
by
\begin{align}
C\geq \max_{\prod_{i\in \mc V}p(x_i)}\min_{d\in \mc D}\min_{\mc
U_d}H( Y_{\mc U_d^c}|X_{\mc U_d^c}, S). \label{EQ: Achievable rate}
\end{align}
\end{theorem}
 The proof of this theorem will be given in Sections
\ref{SEC: layered} and \ref{SEC: arbitrary}.
\begin{remark}
Theorem \ref{TH: AchievableRate} includes the special case of
unicast networks if $|\mc D|=1$.
\end{remark}

\begin{example}[{\cite[Theorem 1]{Dana06}}]
 Consider a network with output symbols $Y_v=\{Y_{u,v}: u\in \mc N_v\}$, where $Y_{u,v}$ is the observation at node
$v$ through the edge $(u,v)$. Thus, the receiving nodes receives a
separate output for each link connected to the node, i.e., has no
interference. Let the output random variables take values from $\mc
Y=\mc X \cup \{e\}$, where the symbol $e$ is the erasure symbol.
Each channel output $Y_{u,v}$ is given by the transmitted signal
$X_u$ with probability $1-\epsilon_{u,v}$ or an erasure symbol $e$
with probability $\epsilon_{u,v}$. Let $S_{u,v,i}$ be a random
variable indicating erasure occurrence across channel $(u,v)\in \mc
E$ at time $i$. If an erasure occurs on link $(u,v)\in \mc E$ at
time $i$, the value of $S_{u,v,i}$ will be one, otherwise zero. Let
$S^n=\{S^n_{u,v}: u\in \mc N_v\}$. If the destination nodes have the
$S^n$ sequence as side information, this channel falls into the
channel model described in Section \ref{SEC: Channel model} since
the output at each relay is a function of the incoming signals and
$S^n$. It can be shown that the cut-set bound is achieved by the
uniform product distribution. Hence, the capacity of this channel is
given by (\ref{EQ: Achievable rate}) with equality.
\end{example}

\subsection{Linear finite field fading networks}
\begin{figure}[!t]
\begin{center}
\leavevmode \epsfxsize=0.27\textwidth \leavevmode
\epsffile{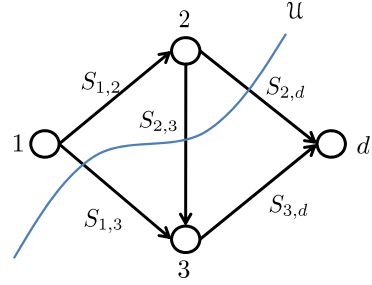}\vspace{-0.2cm}%
\caption{Example of an erasure network. $S_{u,v}$ are erasures
events for links $(u,v)\in \mc E$.}\vspace{-0.4cm}
\label{FIG:erasure network}
\end{center}
\end{figure}
 Consider a finite field $(GF(q))$ network in which each node $v\in \mc V$
observes
\begin{align}
Y_{v}=\sum_{u\in \mc N_v} S_{u,v} X_{u} \label{EQ: ERoutput}
\end{align}
where $Y_{v}$, $X_{u},u\in\mc N_v$, $S_{u,v}, u\in\mc N_v$, are in
$GF(q)$. If we assume that $S_{u,v},\forall (u,v)\in \mc E$ is known
at the destination nodes, this channel falls into the class of
channels in Section \ref{SEC: Channel model}.

 Let $\bb Y_{\mc U^c}$ and $\bb X_{\mc U}$ be vectors of observations in $\bar\partial (\mc
 U^c)$ and input signals in $\partial (\mc U)$, respectively. These are of observations
and input signals of nodes that have an edge passing through the
cut. We define a transfer matrix of an arbitrary cut $\mc U$ as $\bb
G_{\mc U}$ such that it satisfies $\bb Y_{\mc U^c}= \bb G_{\mc U}
\bb X_{\mc U}.$

Thus, the random matrix $\bb G_{\mc U}$ consists of zeros when there
is no connection between the nodes and $S_{u,v}$ if $u \in
\partial (\mc U)$ and $v \in \bar\partial (\mc U^c)$. The column index
represents the sending node index in $\partial (\mc U)$ and row
index represents the receiving node index in $\bar{\partial} (\mc
U^c)$. For the example in Figure \ref{FIG:erasure network}, we have
the expression
\begin{align*}
\underbrace{\left[
  \begin{array}{c}
    Y_3 \\
    Y_d
  \end{array}
\right]}_{\bb Y_{\mc U^c}}=\underbrace{\left[
  \begin{array}{cc}
    S_{1,3} & S_{2,3}\\
    0 & S_{2,d}
  \end{array}
\right]}_{\bb G_{\mc U}}\cdot \underbrace{\left[
  \begin{array}{c}
    X_{1} \\
    X_{2}
  \end{array}
\right]}_{\bb X_{\mc U}}
\end{align*}
for the cut $\mc U=\{1,2\}$.

\begin{theorem} \label{TH: AchievableRate2}
 The multicast capacity of the linear finite field fading network (\ref{EQ: ERoutput}) is
\begin{align*}
C&=\min_{d\in \mc D}\min_{\mc U_d} E[\text{rank} (G_{\mc U_d})]\log
q.
\end{align*}
\end{theorem}
\begin{IEEEproof}
 Proof is omitted due to space limitations.
\end{IEEEproof}
\begin{remark}
For the special case of $S\in\{0,1\}$, Theorem \ref{TH:
AchievableRate2} includes the capacity result for linear finite
field erasure networks with broadcast and interference.
\end{remark}
\section{Proof of Theorem \ref{TH: AchievableRate} for Layered networks} \label{SEC: layered}
\begin{center}
\begin{table*}[ht]
\renewcommand{\arraystretch}{1.2}
{\footnotesize \hfill{}
\caption{Coding strategy of the deterministic diamond network with
state in Fig. \ref{FIG: diamond} } \label{TB: Flow}\hfill
\begin{tabular}{|c|l|l|l|l|l|l|l|}
  \hline
  Block index & Layer 0 & \multicolumn{2}{|c|}{Layer 1 observes}& \multicolumn{2}{|c|}{Layer 1 transmits} & Layer 2 observes & State \\
  \hline
  j & $x_1^n(m_j)$ & $y^n_a(m_j, s^n(j))$ & $y^n_b(m_{j}, s^n(j))$ & $x^n_a(m_{j-1})$ & $x^n_b(m_{j-1})$ & $y_d^n(m_{j-1}, s^n(j))$ & $s^n(j)$ \\
  \hline
  j+1 & $x_1^n(m_{j+1})$ & $y^n_a(m_{j+1}, s^n(j+1))$ & $y^n_b(m_{j+1}, s^n(j+1))$ & $x^n_a(m_{j})$ & $x^n_b(m_{j})$ & $y_d^n(m_{j}, s^n(j+1))$ & $s^n(j+1)$ \\
  \hline
  j+2 & $x_1^n(m_{j+2})$ & $y^n_a(m_{j+2}, s^n(j+2))$ & $y^n_b(m_{j+2}, s^n(j+2))$ & $x^n_a(m_{j+1})$ & $x^n_b(m_{j+1})$ & $y_d^n(m_{j+1}, s^n(j+2))$ & $s^n(j+2)$ \\
  \hline
\end{tabular}}
\hfill{}
\end{table*}
\end{center}

 We begin by showing the achievabililty of Theorem \ref{TH:
 AchievableRate} for layered networks with $\mc D=\{d\}$. The
multicast network is a simple extension of the single destination
network and will be treated later.

 We use a block Markov encoding scheme in which we divide the message $m$
into $K$ parts $m_k, k\in \{1,\ldots, K\}$. We code in $K+L-1$
blocks of length $n$. Message $m_k$ takes values from $[2^{nR}]$ for
all $k$ and the overall rate is given by $\frac{RK}{(K+L-1)}$ which
approaches $R$ as $K\rightarrow \infty$.

 We will use two types of indexing for the inputs, outputs, and state.
We will use $s^n(j)$ to denote the state sequence when message $m_j$
is being sent at the \emph{source} node. For the set of observations
and input sequences at layer $l$ carrying message $m_j$, we will use
the notation
\begin{align}
y^n_{\mc V_l}(m_j)\triangleq \{y_v^n(x^n_{\mc N_v}(m_j), s^n(j+l)):
v\in\mc V_l\} \label{EQ: output}
\end{align}
and
\begin{align}
x^n_{\mc V_l}(m_j)\triangleq \{x_v^n(y_v^n(m_j)): v\in \mc V_l\},
\label{EQ: input}
\end{align}
respectively. For example, $(\ref{EQ: output})$ denotes the set of
observation sequences of the nodes in layer $l$ when $m_j$ is
received. Due to the layered structure of the network and the coding
strategy, which will be explained in the following, the observation
sequences corresponding to the $j$th message at layer $l$ are
functions of $s^n(j+l)$. This will be explained in more detail in
the following.


\emph{Codebook generation:} Fix $p(x_u)$ for all $u\in \mc V-\{d\}$.
Randomly and independently generate $2^{nR}$ sequences $x_1^n(m)$,
$m\in [2^{nR}]$, each according to $\prod_{i=1}^n p(x_{1,i})$. For
each $u\in \mc V-\{1\}$, randomly and independently generate
$x_u^n(y^n_u)$ sequences for each $y^n_u \in \mc Y_u^n$, according
to $\prod_{i=1}^n p(x_{u,i})$.

\emph{Encoding:}
 To send message $m_j$, $j\in\{1,\ldots,K\}$, the encoder sends $x_1^n(m_j)$,
while at each layer $l$, node $v\in \mc V_l$ sends
$x_v^n(y^n_{v}(m_{j-l}))$.

\emph{Decoding:} When the destination receives $y^n_{d}(m_j)$, it
also has $\{s^n(1),\ldots ,s^n(j+L)\}$ from previous observations.
Assuming the previous blocks were decoded with arbitrarily small
error, the receiver declares that a message was sent if it is a
unique index $m_j\in [2^{nR}]$ such that
\begin{align*}
\bigcap_{l=0}^{L-1}\left\{\left(x^n_{\mc V_{l}}(m_j), y^n_{\mc
V_{l+1}}(m_j), s^n(j+l) \right)\in \te\right\}; 
\end{align*}
otherwise an error is declared.

\begin{figure}[!t]
\begin{center}
\leavevmode \epsfxsize=0.5\textwidth \leavevmode
\epsffile{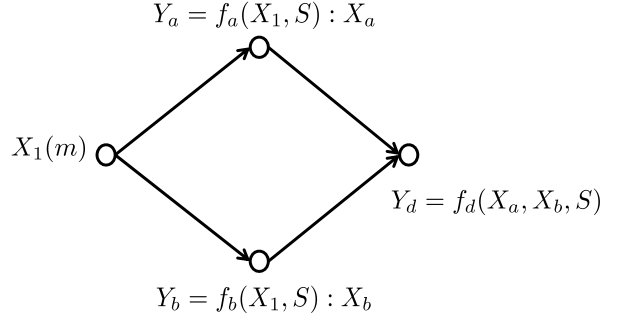}
\caption{Deterministic diamond network with state. }\vspace{-0.4cm}
\label{FIG: diamond}
\end{center}
\end{figure}

From the encoding we can see that there is a $l$ block delay at
layer $l$, $l=\{1,\ldots, L\}$. When the source sends message $m_j$,
the relays in layer 1 send $x^n_{\mc V_1}(m_{j-1})$, the relays in
layer 2 send $x^n_{\mc V_2}(m_{j-2})$ and so on. Accordingly, when
the source sends the $j$th block, received observation sequence of
node $v\in \mc V_l$ is a function of $x^n_{\mc N_v}(m_{j-l-1})$ and
$s^n(j)$, which gives (\ref{EQ: output}). Table \ref{TB: Flow} shows
the coding strategy for a simple diamond network given in Figure
\ref{FIG: diamond}.

 The decoding is a typicality check over an intersection of disjoint
sets. Recall that from (\ref{EQ: output}) and (\ref{EQ: input}), as
message $m_j$ traverses through the network, the message is being
affected by a different state at each layer. Therefore, we require
that all inputs and outputs of that layer and a state (corresponding
to the specific block time) are uniquely jointly typical.

 Before dealing with arbitrarily large networks, we will first give a
proof for a simple diamond network. Consider a diamond network
depicted in Fig. \ref{FIG: diamond} at the top of the next page. The
relay nodes $\{a, b\}$ in layer $1$ receives $Y_a, Y_b$ which are
deterministic functions of $X_1$ and $S$. The destination node in
layer 2 observes $Y_d$, which is a deterministic function of $X_a$,
$X_b$, and $S$.
 Without loss of generality, we will assume that
$m_j=1$ was sent, and show the decoding and probability of error
analysis for the $j$th block. We will omit the message index for
simplicity. There are two types of error events:
\begin{align*}
E_0 \triangleq (A^1_1\cap A^1_2)^c \text{ and } E_1 \triangleq
\bigcup_{m\neq 1} (A_1^{m} \cap A_2^{m})
\end{align*}
where
$$A_1^m \triangleq \left\{\left( X^n_{1}(m), Y^n_{a}(m), Y^n_{b}(m),
S^n(j) \right)\in \te\right\},$$ and
$$A_2^m\triangleq\left\{\left(
X^n_{a}(m), X^n_{b}(m), Y^n_d(1), S^n(j+1) \right)\in \te\right\}.$$
 For the first error event, we have $P(E_0)\rightarrow 0$ as
$n\rightarrow \infty$ by the law of large numbers. We will decompose
$E_1$ into four disjoint events. Let
\begin{align*}
B^m_{\mc Q} &\triangleq \left\{ Y^n_{\mc Q}(m) \neq Y^n_{\mc
Q}(1),Y^n_{\mc Q^c}(m) = Y^n_{\mc Q^c}(1) \right\}
\end{align*}
where $\mc Q \subseteq \{a,b\}$ and $\mc Q^c=\{a,b\}-\mc Q$. We have
four such events since $\{a, b\}$ has four subsets. Then the
probability of $E_1$ is given by
\begin{align}
P(E_1) =& P\left\{\bigcup_{m\neq 1} (A_1^m\cap A_2^m) \right\} \nonumber\\
\leq& \sum_{m\neq 1} P\left\{A_1^m\cap A_2^m\right\} \label{EQ: union1} \\
=& \sum_{m\neq 1} \sum_{\mc Q\subseteq \{a, b\}}P\left\{A_1^m
\cap A_2^m  \cap B^m_{\mc Q}\right\}\label{EQ: partition}
\end{align}
where in (\ref{EQ: union1}) we have used the union bound and
(\ref{EQ: partition}) is from the fact that $B^m_{\mc Q}$ are
partitions that cover the whole set. Thus, we have decomposed $E_1$
into four disjoint events. The event $A_1^m \cap B_{\{a\}}^m$
implies
\begin{align}
\left\{\left( X^n_{1}(m), Y^n_{a}(m), Y^n_{b}(1), S^n(j) \right)\in
\te\right\} \label{EQ: TpSet1}
\end{align}
and $A_2^m \cap B_{\{a\}}^m$ implies
\begin{align}
\left\{\left( X^n_{a}(m), X^n_{b}(1), Y^n_d(1), S^n(j+1) \right)\in
\te\right\} \label{EQ: TpSet2}
\end{align}
since $X^n_b(m)=X^n_{b}(Y_b^n(m))$. 
Since (\ref{EQ: TpSet1}) and (\ref{EQ: TpSet2}) are independent
events, we have
\begin{align}
&P\{A_1^m \cap A_2^m\cap B_{\{a\}}^m\}\nonumber\\
&\leq 2^{-n(I(X_1, Y_a; Y_b|S)-3\epsilon)}2^{-n(I(X_a; Y_d|X_b,
S)-3\epsilon)}\nonumber\\
&= 2^{-n(H(Y_b, Y_d|S, X_b)-6\epsilon)} \label{EQ: bound1}
\end{align}
where in the last step we have used the Markov structure of the
network.
Similar to the previous steps, we can bound the other events by
\begin{align}
P\{A_1^m \cap A_2^m\cap B_{\{b\}}^m\}&\leq 2^{-n(H(Y_a, Y_d|S, X_a)-6\epsilon)}\label{EQ: bound2},\\
P\{A_1^m \cap A_2^m\cap B_{\phi}^m\} &\leq2^{-n(H(Y_a, Y_b|S, X_a,
X_b)-6\epsilon)}\label{EQ: bound3},
\end{align}
and
\begin{align}
P\{A_1^m \cap A_2^m\cap
B_{\{a,b\}}^m\}\leq2^{-n(H(Y_d|S)-3\epsilon)}. \label{EQ: bound4}
\end{align}
Combining (\ref{EQ: partition}), (\ref{EQ: bound1}), (\ref{EQ:
bound2}), (\ref{EQ: bound3}), and (\ref{EQ: bound4}), we get
$P(E_1)\rightarrow 0$ as $n\rightarrow \infty$ if
\begin{align}
R< \min \left\{
  \begin{array}{c}
    H(Y_a, Y_b, Y_d|S, X_a, X_b)-6\epsilon,\\ H(Y_d|S)-3\epsilon,\\
    H(Y_b, Y_d|S, X_b)-6\epsilon,\\ H(Y_a, Y_d|S, X_a)-6\epsilon
  \end{array}
\right\},
\end{align}
which concludes the proof for the diamond network.

 We now move on to the proof of Theorem \ref{TH: AchievableRate} for
general layered networks. We will show the proof for decoding
message $m_j$, and define similar events as in the diamond network
to lead us through the proof. As before, we omit the message index
for simplicity. Let
\begin{align}
A_l^m \triangleq \left\{\left(X^n_{\mc V_{l}}(m), Y^n_{\mc
V_{l+1}}(m), S^n(j+l) \right)\in \te\right\} \label{EQ: Al}.
\end{align}
 Notice that we are abusing notation for the destination
observation in (\ref{EQ: Al}). For $A_{L-1}^m$, $Y^n_{\mc V_{L}}(m)$
should be $Y^n_d$, which is the given observation at the destination
and is not \emph{tested} for typicality.

 Assuming $m=1$ was sent, we have two sources of error:
$$E_0 \triangleq
\left(\bigcap_{l=0}^{L-1}A^1_l\right)^c \text{ and } E_1\triangleq
\bigcup_{m\neq 1}\bigcap_{l=0}^{L-1}A_l^m.$$ The error event
$P(E_0)\rightarrow 0$ as $n \rightarrow \infty$. As we did in the
diamond network case we will decompose the error event $E_1$ with
each $B^m_{\mc Q}$, $\mc Q\subseteq \mc R$. The probability of $E_1$
is given by
\begin{align}
P(E_1)=& P\left\{\bigcup_{m\neq 1} \bigcap_{l=0}^{L-1}A_l^m
\right\} \nonumber\\
\leq& \sum_{m\neq 1}P\left\{\bigcap_{l=0}^{L-1}A_l^m \right\} \nonumber\\
=& \sum_{m\neq 1}\sum_{\mc Q \subseteq \mc R} P\left\{
\bigcap_{l=0}^{L-1}A_l^m\cap B^m_{\mc Q}\right\} \label{EQ: final1}
\end{align}
where the inequality is due to the union bound and the last step is
due to partitioning the events. The event $A_{l}^m \cap B^m_{\mc Q}$
implies
\begin{small}
\begin{align*}
\left\{\left(X^n_{\mc Q_{l}}(m), X^n_{\mc Q^c_{l}}(1), Y^n_{\mc
Q_{l+1}}(m), Y^n_{\mc Q_{l+1}^c }(1), S^n(j+l) \right)\in
\te\right\}
\end{align*}
\end{small}
where $\mc Q_l=\mc V_l \cap \mc Q$ and $\mc Q_l^c=\mc V_l - \mc
Q_l$. Then,
\begin{align}
P\left\{\bigcap_{l=0}^{L-1}A_l^m\cap B^m_{\mc Q}\right\}&\leq
\prod_{l=0}^{L-1}2^{-n(I(X_{\mc Q_l};
Y_{\mc Q_{l+1}^c}|X_{\mc Q_l^c}, S)-3\epsilon)} \nn\\
&=\prod_{l=0}^{L-1} 2^{-n(H(Y_{\mc Q_{l+1}^c}|X_{\mc Q_l^c},
S)-3\epsilon)}. \label{EQ: final2}
\end{align}
From (\ref{EQ: final1}) and (\ref{EQ: final2}) we get
\begin{align}
P(E_1) 
\leq&\sum_{m\neq 1}\sum_{\mc Q \subseteq \mc R}
\prod_{l=0}^{L-1}2^{-n(H(Y_{\mc Q_{l+1}^c}|X_{\mc Q_l^c},
S)-3\epsilon)} \nn \\
= &\sum_{m\neq 1}\sum_{\mc Q \subseteq \mc R}
2^{-n\sum_{l=0}^{L-1}(H(Y_{\mc Q_{l+1}^c}|X_{\mc Q_l^c},
S)-3\epsilon)} \nn \\
\leq &\sum_{\mc Q \subseteq \mc R}
2^{nR}2^{-n\sum_{l=0}^{L-1}(H(Y_{\mc Q_{l+1}^c}|X_{\mc Q_l^c},
S)-3\epsilon)} \nn \\
= &\sum_{\mc Q \subseteq \mc R} 2^{nR}2^{-n(H(Y_{\mc U^c}|X_{\mc
U^c}, S)-\epsilon')} \nn
\end{align}
where $\epsilon'=3L\epsilon$ and $\mc U^c=\{\mc Q^c, d\}$ which
gives a cut in the network. Thus, $P(E_1) \rightarrow 0$ as $n
\rightarrow \infty$ if
\begin{align*}
R<\min_{\mc U} H(Y_{\mc U^c}|X_{\mc U^c}, S)-\epsilon',
\end{align*}
which proves Theorem \ref{TH: AchievableRate} for layered networks
with a single destination.
\begin{remark}
 Consider a semi-deterministic layered network $\mc G=(\mc V, \mc E)$ where each node
$v\in \mc V-\{d\}$ observes $Y_v=f_v(X_{\mc N_v}, Y_d)$ and the
final destination gets $Y_d \sim p(y_d|x_{\mc N_d})$, i.e., a
stochastic output. Using the coding scheme above we can show that
all rates $R$ that satisfies
$$R< \max_{\prod_{i\in \mc V}p(x_i)}\min_{\mc
U}I( X_{\mc U}; Y_{\mc U^c}|X_{\mc U^c})$$ are achievable for
unicast.
\end{remark}

 For the multicast scenario we declare an error if any of the nodes in $\mc D$ makes an
error. Using the union bound and the same line of proof as in
Section \ref{SEC: layered} for each $d\in \mc D$, we can show that
the probability of error is arbitrarily small for sufficiently large
$n$ if
\begin{align*}
R< \max_{\prod_{i\in \mc V}p(x_i)}\min_{d\in \mc D} \min_{\mc U_d}
H(Y_{\mc U_d^c}|X_{\mc U_d^c}, S).
\end{align*}

\section{Arbitrary networks} \label{SEC: arbitrary}
 For extending the layered network result to arbitrary networks
we use the same line of proof as done in \cite{Avestimehr07} that
unfolds $\mc G$ into a time-extended network. We will just give an
outline of the proof. For more details on unfolding $\mc G$, we
refer to \cite{Avestimehr07} due to space limitations. Given an
arbitrary network $\mc G$, we unfold the original network over $T$
stages to get a layered network $\bar{\mc G}$. Using the coding
scheme for the unfolded layered network, we can achieve
\begin{align}
R< \frac{1}{T}\max_{\prod_{i\in \mc V}p(x_i)}\min_{\bar{\mc U}}
H(Y_{\bar{\mc U}^c}|X_{\bar{\mc U}^c}, S) \label{EQ: wigtranslate}
\end{align}
where $\bar{\mc U}$ is a cut in the unfolded network. We normalize
the right hand side by $T$ since the network gives at most $T$
duplicate paths of the original network. Using Lemma 6.2 in
\cite{Avestimehr07} (by including a state random variable in the
conditional entropies) we have the relation
\begin{align}
(T+N-1) \min_{\mc U} H(Y_{\mc U^c}|X_{\mc U^c}, S)\leq H(Y_{\bar{\mc
U}^c}|X_{\bar{\mc U}^c}, S) \label{EQ: wiggling}
\end{align}
where $N=2^{|\mc V|-2}$. We also have for any distribution,
\begin{align}
 \min_{\bar{\mc U}} H(Y_{\bar{\mc U}^c}|X_{\bar{\mc U}^c}, S)\leq  T \min_{\mc U} H(Y_{\mc U^c}|X_{\mc
U^c}, S), \label{EQ: wigupper}
\end{align}
since the right hand side corresponds to taking the minimum over
only steady cuts (subset of all possible cuts). Combining (\ref{EQ:
wiggling}) with (\ref{EQ: wigupper}) we have
\begin{align*}
\lim_{T \rightarrow \infty} \frac{1}{T}&\max_{\prod_{v\in \mc
V}p(x_v)}\min_{\bar{\mc U}} H(Y_{\bar{\mc U}^c}|X_{\bar{\mc U}^c},
S) \\
&\leq  \max_{\prod_{v\in \mc V}p(x_v)} \min_{\mc U} H(Y_{\mc
U^c}|X_{\mc U^c}, S).
\end{align*}
Finally, with the relations (\ref{EQ: wigtranslate}) and (\ref{EQ:
wigupper}), we can show that rates arbitrary close to the right hand
side of (\ref{EQ: Achievable rate}) are achievable for sufficiently
large $T$.
\section*{Acknowledgment}
The work of Sung Hoon Lim and Sae-Young Chung is partially supported
by the MKE, Korea, under the ITRC support program supervised by the
IITA (IITA-2009-C1090-0902-0005), and the work of Young-Han Kim is
partially supported by the National Science Foundation CAREER award
CCF-0747111.

\bibliographystyle{IEEEtran}
\bibliography{myref}

\begin{thebibliography}{1}
\providecommand{\url}[1]{#1}
\csname url@samestyle\endcsname
\providecommand{\newblock}{\relax}
\providecommand{\bibinfo}[2]{#2}
\providecommand{\BIBentrySTDinterwordspacing}{\spaceskip=0pt\relax}
\providecommand{\BIBentryALTinterwordstretchfactor}{4}
\providecommand{\BIBentryALTinterwordspacing}{\spaceskip=\fontdimen2\font plus
\BIBentryALTinterwordstretchfactor\fontdimen3\font minus
  \fontdimen4\font\relax}
\providecommand{\BIBforeignlanguage}[2]{{%
\expandafter\ifx\csname l@#1\endcsname\relax
\typeout{** WARNING: IEEEtran.bst: No hyphenation pattern has been}%
\typeout{** loaded for the language `#1'. Using the pattern for}%
\typeout{** the default language instead.}%
\else
\language=\csname l@#1\endcsname
\fi
#2}}
\providecommand{\BIBdecl}{\relax}
\BIBdecl

\bibitem{Ahlswede00}
R.~Ahlswede, N.~Cai, S.-Y. Li, and R.~Yeung, ``Network information flow,''
  \emph{IEEE Trans. Inf. Theory}, vol.~46, no.~4, pp. 1204--1216, Jul 2000.

\bibitem{Ratnakar06}
N.~Ratnakar and G.~Kramer, ``The multicast capacity of deterministic relay
  networks with no interference,'' \emph{IEEE Trans. Inf. Theory}, vol.~52,
  no.~6, pp. 2425--2432, June 2006.

\bibitem{Avestimehr07}
A.~Avestimehr, S.~N. Diggavi, and D.~Tse, ``Wireless network information
  flow,'' in \emph{Proc. Forty-Fifth Annual Allerton Conf. Commun., Contr.
  Comput.}, Monticello, IL, Sept. 2007.

\bibitem{Avestimehr08}
------, ``Approximate capacity of {G}aussian relay networks,'' in \emph{Proc.
  IEEE Int. Symp. Information Theory}, Toronto, Ontario, Canada, 2008, pp.
  474--478.

\bibitem{Dana06}
A.~Dana, R.~Gowaikar, R.~Palanki, B.~Hassibi, and M.~Effros, ``Capacity of
  wireless erasure networks,'' \emph{IEEE Trans. Inf. Theory}, vol.~52, no.~3,
  pp. 789--804, March 2006.

\bibitem{Smith07}
B.~Smith and S.~Vishwanath, ``Unicast transmission over multiple access erasure
  networks: Capacity and duality,'' in \emph{Proc. IEEE Information Theory
  Workshop}, Tahoe City, California, Sept. 2007, pp. 331--336.

\bibitem{Orlitsky01}
A.~Orlitsky and J.~Roche, ``Coding for computing,'' \emph{IEEE Trans. Inf.
  Theory}, vol.~47, no.~3, pp. 903--917, Mar 2001.

\bibitem{Cover06}
T.~M. Cover and J.~A. Thomas, \emph{Elements of Information Theory},
  2nd~ed.\hskip 1em plus 0.5em minus 0.4em\relax New York: Wiley, 2006.

\end{thebibliography}
\end{document}